\documentclass[letter]{aa}  

\usepackage{graphicx}
\usepackage{txfonts}
\usepackage[utf8]{inputenc}
\usepackage{amsfonts}
\usepackage{amsmath}
\usepackage{bm}
\usepackage{xcolor}
\usepackage{color}
\usepackage{amssymb}
\usepackage{natbib}
\usepackage{xspace}
\usepackage{ulem}
\usepackage{orcidlink}

\begin{document} 

\title{Energetic particle-mediated interplanetary shocks observed by Solar Orbiter}
\titlerunning{Energetic particles-mediated IP shocks}
\author{D. Trotta\orcidlink{0000-0002-0608-8897}\inst{1}
\and 
D. Lario\orcidlink{0000-0002-3176-8704}\inst{2}
\and
B. Reville\orcidlink{0000-0002-3778-1432}\inst{3}
\and
S. Raptis\orcidlink{0000-0002-4381-3197}\inst{4}
\and
O. Pezzi\orcidlink{0000-0002-7638-1706}\inst{5}
\and
H. Hietala\orcidlink{0000-0002-3039-1255}\inst{6}
\and
P. Mostafavi\orcidlink{0000-0002-3808-3580}\inst{4}
\and
J. Giacalone\orcidlink{0000-0002-0850-4233}\inst{7}
\and
R. F. Wimmer-Schweingruber\orcidlink{0000-0002-7388-173X}\inst{8}
\and
P. Kuehl\orcidlink{0000-0002-3758-9272}\inst{8}
\and
A. Kollhoff\orcidlink{0000-0002-9471-5132}\inst{8}
\and
D. Turner\inst{4}
\and
D. Burgess\inst{6}}

\authorrunning{D. Trotta et al.}
\institute{European Space Agency (ESA), European Space Astronomy Centre (ESAC), Camino Bajo del Castillo s/n, 28692 Villanueva de la Cañada, Madrid, Spain \email{domenico.trotta@esa.int}
\and  
Heliospheric Physics Laboratory, Heliophysics Division, NASA Goddard Space Flight Center, Greenbelt, MD 20771, USA 
\and
Max-Planck-Institut für Kernphysik, Heidelberg, Germany
\and
Johns Hopkins University Applied Physics Laboratory, Laurel, MD, USA 
\and
Istituto per la Scienza e Tecnologia dei Plasmi (ISTP), Consiglio Nazionale delle Ricerche, I-70126 Bari, Italy 
\and
Department of Physics and Astronomy, Queen Mary University of London, Mile End Road,
London E1 4NS, UK
\and
Lunar and Planetary Laboratory, University of Arizona, Tucson, USA  
\and
Institute of Experimental and Applied Physics, Kiel University, Leibnizstrasse 11, D-24118 Kiel, Germany 
}

\date{\today}

\abstract
  {In collisionless shocks, energetic particles can carry sufficient pressure to modify the upstream plasma and the shock structure itself, a regime often invoked in theories of cosmic-ray acceleration but rarely observed in the heliosphere.}
  {We find and characterize {interplanetary} IP shocks where energetic particles dynamically dominate the upstream pressure.}
  {We analyse IP shocks observed by Solar Orbiter inside 1 au and compute the energetic particle pressure $P_{EP}$ from proton measurements above 10\,keV, comparing it with the upstream thermal $P_{Th}$ and magnetic $P_{B}$ pressures.}
  {We identify four shocks for which $P_{EP} \geq P_{Th} + P_B $. These events correspond to strong and fast shocks in the high-Mach-number tail of the Solar Orbiter shock population. In several cases the $P_{EP}$ increase coincides with a decreasing upstream bulk flow speed in the shock frame, and the resulting particle-mediated foreshocks extend up to $\sim10^5$ {ion inertial lengths} $d_i$. The extent of such energetic particle dominated region depends on shock geometry.}
  {These observations provide evidence that accelerated particles can dynamically modify interplanetary shocks. They highlight the importance of the coupling between energetic particles, upstream fluctuations, and shock structure for understanding particle acceleration at collisionless shocks.}
  

\maketitle

\section{Introduction}
\label{sect:intro}

Energetic particles are ubiquitous in astrophysical environments, yet the mechanisms responsible for their production have been debated since the seminal works of Fermi in the 1950s~\citep{Fermi1949,Fermi1954}. {Collisionless shocks, sharp transitions from super-sonic/super-Alfvénic to sub-sonic/sub-Alfvénic flows that convert directed upstream kinetic energy into downstream thermal and magnetic energy, can channel a significant fraction of this energy into non-thermal particles and are therefore among the most efficient particle accelerators in astrophysical plasmas.}~\citep{BaloghTreumann2013,Bykov2019,Giuffrida2022}.

Shocks in the heliosphere provide a unique opportunity to study particle acceleration in situ, unlike shocks in distant astrophysical systems where only indirect diagnostics are possible~\citep{StoneTsurutani1985}. These observations have provided key tests for particle acceleration theories~\citep[e.g.,][]{Schwartz2022,Amano2020}. However, whether insights from heliospheric shocks can be directly extrapolated to astrophysical systems remains debated, because they occupy different regions of parameter space (see, e.g., \citep{2013SSRv..178..633G}). For example, Coronal Mass Ejection (CME)-driven interplanetary (IP) shocks typically have lower Mach numbers than many astrophysical shocks~\citep[e.g.,][]{Burgess2015,PerezAlanis2023}.

A particularly important yet poorly explored regime is the energetic particle (or cosmic-ray) dominated regime, in which energetic particle pressure $P_{EP}$ exceeds the thermal $P_{Th}$ and magnetic field $P_{B}$ pressures upstream of the shocks~\citep{Drury1983}. Simulations predict that gradients in $P_{EP}$ may form an upstream precursor that decelerates and modifies the shock, producing compression ratios exceeding the canonical value of four for strong shocks~\citep{RevilleBell2013,Haggerty2020}. These effects have important implications for energetic particle production and transport~\citep{JonesEllison1991}.

Observational hints of this regime in the heliosphere have been reported previously near Earth~\citep{Lario2002} and at the heliospheric termination shock~\citep{Florinski2004}. \citet{Terasawa2006} analysed two strong IP shocks observed by Geotail and found signatures consistent with the energetic particle mediated regime, although the energetic particle pressure was not quantified. Later, \citet{Lario2015} identified periods when energetic proton pressure exceeded both magnetic and thermal pressures upstream of IP shocks using STEREO observations, but the limited energy coverage limited the characterisation of these precursors.

The Solar Orbiter mission~\citep{Muller2020} provides new opportunities to investigate these processes thanks to its advanced energetic particle instrumentation~\citep{RodriguezPacheco2020,Wimmer2021}. With more than 150 shock events observed since launch until end of 2025~\citep{Trotta2025b}, it enables detailed studies of shock-accelerated particles with unprecedented coverage~\citep[e.g.,][]{Trotta2023c,Yang2024}.

In this work, we identify and characterise four energetic particle-mediated shocks, thereby providing the first detailed description of energetic particle-mediated shocks using in-situ observations. Full details of the instrumentation used and the energetic particle and IP context where these shocks were observed are given in Appendices~\ref{appendix:data} and ~\ref{appendix:seeds}, respectively.
 \begin{figure*}  
   \includegraphics[width=0.99\textwidth]{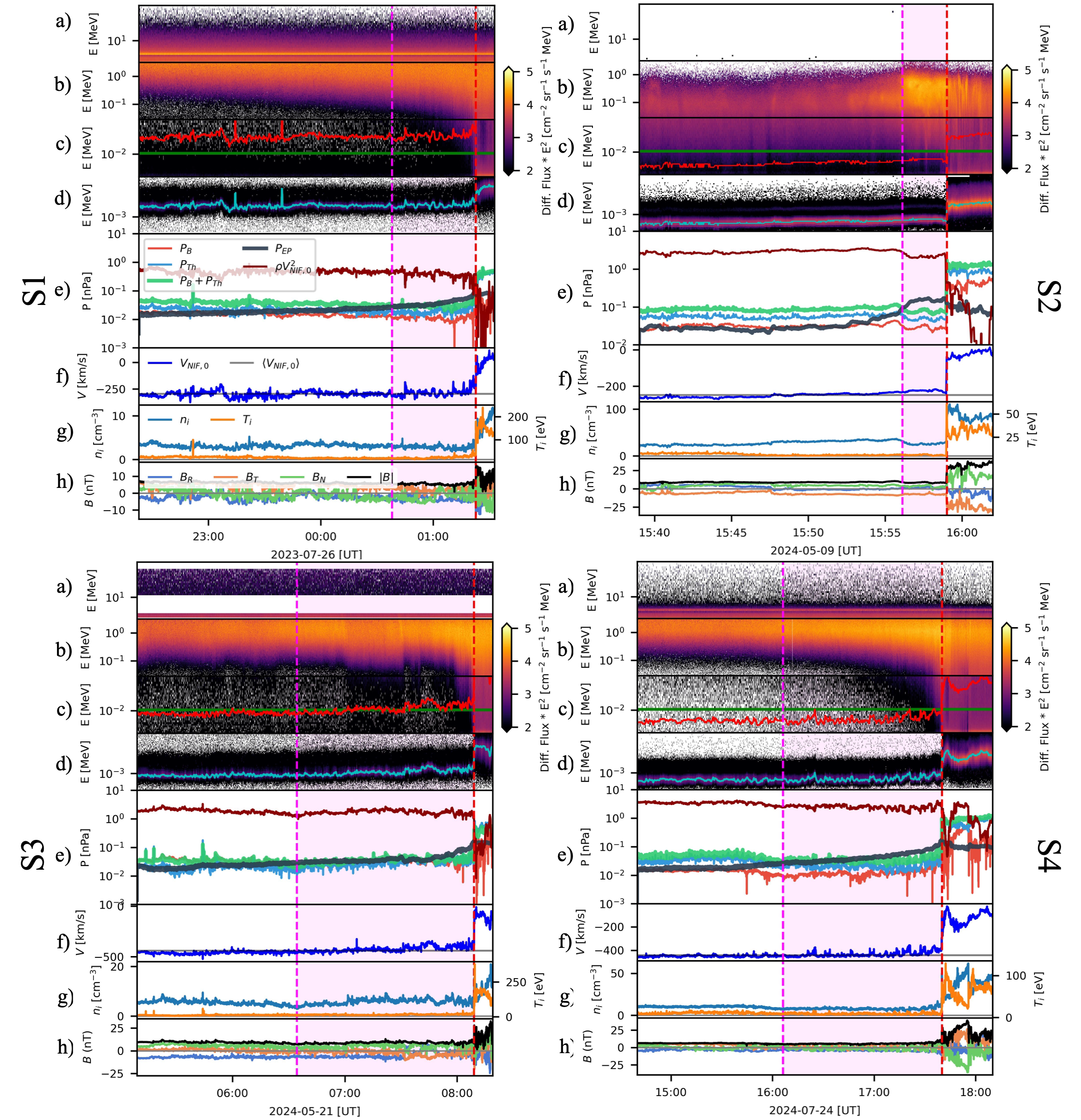}
   \caption{Overview of the four energetic particle-mediated shocks. a)–d): Differential energy flux measured by the sunward apertures of HET/EPT (a,b), pixel-averaged STEP (c), and angle-averaged SWA-PAS (d). The cyan line marks the peak of the thermal proton distribution $E_{bulk}$, the red line shows $10 E_{bulk}$, and the green line indicates the lower integration limit $E_1$ used in the energetic particle pressure calculation. e): Evolution of the pressure contributions: magnetic pressure $P_B$ (red), thermal pressure $P_{Th}$ (blue), energetic particle pressure $P_{EP}$ (black), and $P_{Th}+P_{EP}$ (green) and ram pressure $\rho V_{NIF,0}^2$ along the shock normal in the shock rest frame (dark red). Magenta shading marks intervals where $P_{EP} \geq P_B + P_{Th}$.  f): Plasma velocity along the shock normal in the shock rest frame (blue), with the grey line indicating the upstream average. Panels g) and h) show the ion density and temperature, and the magnetic field magnitude and components, respectively.}
              \label{fig:overview}%
\end{figure*}

\begin{table*}
\caption{Energetic particle mediated shocks and their parameters. For full details about how these are computed, see Appendix~\ref{appendix:shock_params_frame}.}             
\label{table:shocks}      
\centering          
\begin{tabular}{l c | c c c c c c c c c c }     
\hline
Name & Shock Time [UTC] &$\mathrm{r_G}$ & $\mathrm{r_B}$ &Normal (RTN) & $\mathrm{v_{sh}}$  [km/s] & $\mathrm{M_A}$ & $\mathrm{M_{fms}}$ & $\theta_{Bn}$ [$^\circ$] & $ {\tau_{EP}}$ [min] & $L_{EP}$ [$d_i$]  \\ 
\hline                    
S1 & 2023-07-26 01:22:46  & 2.1 & 2.1 & [0.88, -0.18,  0.44] & 997 & 6.3 & 4.3 & 48 & 44 & 19500 \\ 
S2 & 2024-05-09 15:58:59  & 3.5 & 3.7 & [0.97, -0.17,  0.13] & 637 & 7.7 & 5.8 & 72 & 3  &  2500 \\ 
S3 & 2024-05-21 08:09:05  & 2.0 & 2.2 & [0.91,  0.21, -0.36] & 990 & 6.5 & 5.5 & 37 & 95 & 59000 \\ 
S4 & 2024-07-24 17:40:01  & 2.1 & 2.2 & [0.98, -0.04,  0.20] & 820 & 8.1 & 6.9 & 40 & 93 & 56500 \\ 
\hline                
\end{tabular}
\end{table*}

 \begin{figure}  
    \centering
   \includegraphics[width=0.45\textwidth]{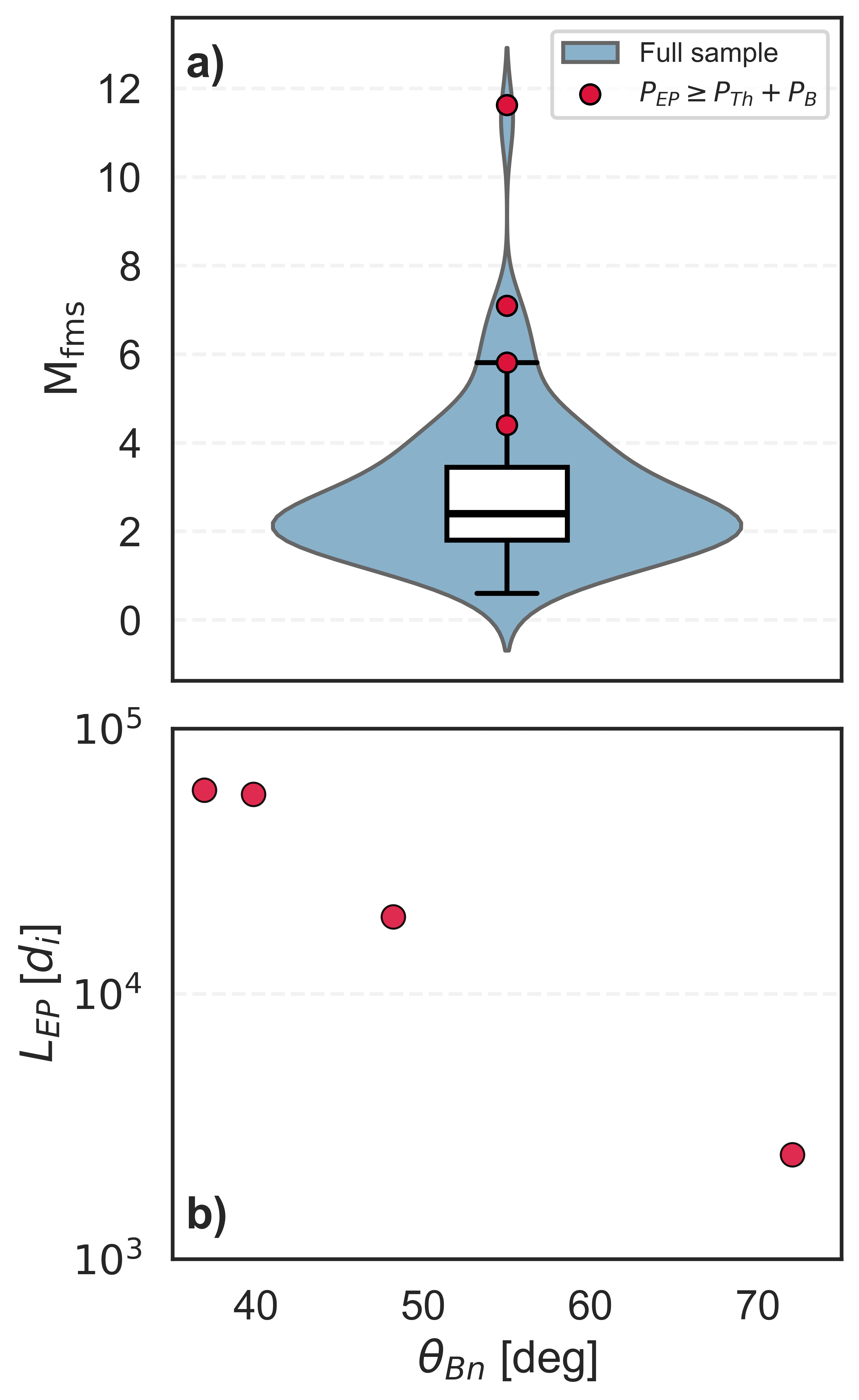}
   \caption{a) Distribution of the fast magnetosonic Mach number $\mathrm{M_{fms}}$ for the Solar Orbiter shocks. The violin plot shows the full distribution, with the box indicating the interquartile range and median. The four shocks identified in this work with $P_{EP} \geq P_B + P_{Th}$ are highlighted in red. b) Upstream precursor length $L_{EP}$ as a function of the shock normal angle $\theta_{Bn}$ for the four energetic particle-mediated shocks.}
              \label{fig:fig2_stats}%
\end{figure}

\section{Results and Discussion}\label{sec:results}

For each event in the Solar Orbiter shock sample, extending that reported in \citet{Trotta2025b} to 152 events, we computed $P_{EP}$ 5 hours before and 1 hour after the shock passage using the expression $P_{EP} \equiv (4\pi/3) (2m)^{1/2} \int_{E_1}^{E_2} \sqrt{E} j(E) dE$, where $E$ is the proton kinetic energy, $j(E)$ the proton differential flux, and $m$ the proton mass. We use $E_1 =$ 10 keV and $E_2 =$ 100 MeV as the lower and upper integration bounds~\citep[see][]{Russell2013,Lario2015}. $P_{EP}$ is then compared to $P_B + P_{Th}$, where $P_B \equiv B^2/2\mu_0$ is the magnetic pressure ($B$ and $\mu_0$ are the magnetic field magnitude and magnetic permeability) and $P_{Th} = 2 n_p k_B T_p$ is the thermal pressure ($n_{p}$ and $T_{p}$ are the solar wind proton density and temperature, and $k_{B}$ the Boltzmann constant). Here we assume the electron temperature $T_e \simeq T_p$~\citep[see][]{Shi2023}. We identify periods where $P_{EP}\geq P_{B}+ P_{Th}$ upstream of the shocks, yielding four events in this regime.

An overview of these shocks is shown in Figure~\ref{fig:overview}. For each event, panels a)–d) display the differential energy flux measured by the sunward apertures of HET/EPT (a,b), pixel-averaged STEP (c), and angle-averaged SWA-PAS (d). The cyan line in panel d) marks the peak of the thermal proton distribution, $E_{bulk}$. The red line in panel c) shows $10\times E_{bulk}$, while the green line indicates the lowest energy $E_1$ used in the $P_{EP}$ computation. In most cases $E_1$ lies above $10\times E_{bulk}$, confirming that the pressure estimate is dominated by suprathermal and energetic particles. The only exception is S1, where $E_1$ falls below this threshold; however, repeating the calculation with a higher $E_1$ does not change the results.

Panels e) show the evolution of the pressure contributions during each event, with $P_B$, $P_{Th}$, $P_{EP}$, and $P_{Th}+P_{B}$ shown as red, blue, black, and green lines, respectively. The magenta shaded regions mark intervals where $P_{EP} \geq P_B + P_{Th}$. We include in panels e) the ram pressure along the shock normal $P_{ram}$, computed in the normal incidence shock frame (NIF; see Appendix~\ref{appendix:shock_params_frame}). In several cases, decreases in $P_{ram}$ coincide with the largest increases in $P_{EP}$.

Panels f) show the plasma velocity along the shock normal in the shock rest frame (blue),with the horizontal grey line indicating its upstream average value. In three of the four events the upstream velocity increases prior to the shock, consistent with the action of an energetic particle pressure precursor; this signature is weaker for S1. Panels g) and h) show $n_{p}$ and $T_{p}$ (g), and the magnetic field magnitude and components (h).

Figure 1 shows that the periods with $P_{EP}\geq P_{Th}+P_{B}$ occur under a range of upstream conditions, from the short-lived event of S2 in a relatively quiet solar wind to the prolonged case of S4 embedded in a highly structured upstream medium. The properties of the four shocks are summarised in Table~\ref{table:shocks}. Full details on the parameters and their determination are provided in Appendix~\ref{appendix:shock_params_frame}. Together with typical shock parameters, we report $\tau_{EP}$, the duration for which $P_{EP} \geq P_{Th} + P_B$ upstream of the shock, and $L_{EP}$, obtained by converting this duration into a spatial scale using the shock speed and expressing it in units of the ion inertial length $d_i$. S1,2,3 and 4 occur at heliocentric distances of .92, .68, .72 and .92 AU, respectively.

All events correspond to fast and strong shocks, with shock speeds between $\sim 600$ and $1000$ km s$^{-1}$ and Alfvénic and fast magnetosonic Mach numbers in the ranges $\mathrm{M_A}\sim6$–8 and $\mathrm{M_{fms}}\sim4$–7. The shocks span a range of geometries, from quasi-parallel to quasi-perpendicular ($\theta_{Bn}=35^\circ$–$72^\circ$), indicating that the energetic particle-mediated regime is not restricted to a specific shock orientation. The duration of the energetic particle precursor $\tau_{EP}$ varies widely, from only a few minutes for S2 to more than 100 minutes for S3 and 4, corresponding to precursor scales of $\sim10^3$ – $10^5\,d_i$. The peak value of $P_{EP}/(P_{B}+P_{th})$ is of 2.5, 2.3, 3 and 3.1 for S1 to 4, respectively.

These results indicate that extended energetic-particle dominated precursors are generated upstream of strong fast shocks. From this point of view, our results show that energetic particle-mediated shocks require parameters at the extreme of the IP-shock parameter distributions~\citep{Kilpua2015_shocks,PerezAlanis2023}. This finding is consistent with previous observations~\citep{Russell2013,Terasawa2006,Lario2015} and with theoretical expectations that stronger shocks are more efficient at accelerating particles and can therefore more readily reach the energetic particle mediated regime~\citep[e.g.,][]{Bell2013}.

To further quantify how unusual these shocks are, we compare their fast magnetosonic Mach numbers with the full Solar Orbiter shock sample. The violin plot in Figure~\ref{fig:fig2_stats}a) shows the distribution of $\mathrm{M_{fms}}$ for all shocks, with the four energetic particle-mediated events highlighted. All four lie in the high-Mach-number tail of the distribution, demonstrating that the energetic particle-mediated regime occurs only for the strongest IP shocks, with an occurrence of about 3\% over the 152 events analyzed. Similar results are obtained when when using M$_{A}$ or the shock speed.

To investigate which shock properties control the development of the energetic particle precursor, we explored correlations between the precursor length $L_{EP}$ and the main shock parameters, including shock speed, Mach numbers, and shock geometry. Among the quantities examined, the shock normal angle $\theta_{Bn}$ shows the clearest relationship with the precursor extent. Figure~\ref{fig:fig2_stats}b) illustrates this trend by showing the precursor length $L_{EP}$ as a function of $\theta_{Bn}$. This behavior is consistent with the role of shock geometry in controlling particle transport upstream of the shock. Quasi--parallel shocks (smaller $\theta_{Bn}$) allow particles to escape farther upstream, producing extended precursors, whereas quasi--perpendicular shocks tend to confine particles closer to the shock~\citep{Lario2019}. The observed variation of $L_{EP}$ with $\theta_{Bn}$ is therefore consistent with theoretical expectations for energetic particle transport at collisionless shocks~\citep{RevilleBell2013}.

We also note that three of the four shocks (all except the quasi-perpendicular shock 2) occur after large solar energetic particle (SEP) events (Appendix~\ref{appendix:seeds}), providing a substantial ``seed'' population of energetic particles upstream of the shock~\citep[see][]{Kahler2000,Wijsen23}. In addition, the two most quasi-parallel shocks in our sample exhibit signatures of a wave foreshock upstream of the shock (see Appendix~\ref{appendix:foreshock}).

\section{Conclusions}\label{sec:conclusions}
We identified four interplanetary shocks in the Solar Orbiter sample for which the upstream energetic particle pressure exceeds the combined thermal and magnetic pressures, placing them in the energetic particle-mediated regime. These events are associated with strong and fast shocks that lie in the high-Mach-number tail of the Solar Orbiter shock population. In several cases, the increase in energetic particle pressure coincides with a decrease in ram pressure and a decrease of upstream bulk flow speed in the shock frame. The spatial extent of these energetic particle-mediated foreshocks reaches $\sim10^3$–$10^5,d_i$ and shows a dependence on shock geometry, with more quasi-parallel shocks producing more extended energetic particle-dominated foreshocks. We also note that three of the four shocks follow SEP events, suggesting that a substantial seed population may play a role in the development of this regime. {The highly dynamic energetic-particle signatures of these four events provide a unique opportunity to probe particle–shock–ambient-plasma coupling~\citep{Wimmer2026}}.

These observations provide direct evidence that accelerated particles can dynamically modify the upstream medium of IP shocks, linking energetic particle pressure, upstream fluctuations, and shock structure. Upstream particle-driven modifications are also observed at Earth's bow shock, although they are typically associated with lower-energy particles and occur in a stationary shock environment~\citep{Johlander2021,Lalti2022}. Recent comparisons between planetary and interplanetary compressive foreshocks further highlight these connections~\citep{Raptis2026}. A direct comparison with the energetic particle-mediated interplanetary shocks identified here will be the subject of future work.

\bibliographystyle{aa}
\bibliography{bibby_abbrev}

\begin{appendix}
\section{Data Products}\label{appendix:data}
We use the full in situ Solar Orbiter payload. Magnetic field data are obtained from the flux-gate magnetometer~\citep[MAG;][]{Horbury2020}, operating at 64 vectors s$^{-1}$ in burst mode. Ion energy fluxes and plasma moments are provided by the Proton Alpha Sensor (PAS) of the Solar Wind Analyzer (SWA) suite~\citep{Owen2020} at a cadence of 4 s. For S1, the electron density is also estimated from the spacecraft potential measured by the Radio and Plasma Waves instrument~\citep[RPW;][]{Maksimovic2020}, available at a temporal resolution of 0.01 s. We used this product to check the consistency with the PAS--measured density.

Energetic particles are measured by the Energetic Particle Detector (EPD) suite~\citep{RodriguezPacheco2020}. We use the SupraThermal Electrons and Protons (STEP) sensor, which measures ions from a few keV up to $\sim80$ keV using 15 sunward-pointing pixels. We use higher-energy ions measured by the sunward apertures of the Electron Proton Telescope (EPT), covering energies from 25 keV to 6.4 MeV, and of the High Energy Telescope (HET), covering energies from 6.8 to above 100 MeV nuc$^{-1}$ depending on the ion species, from which in this work we use only proton measurements. All energetic particle datasets are used at 5 s resolution. The fields of view of STEP, EPT, and HET partially overlap and have angular widths of approximately $30^\circ$.

\section{Shock parameters and frames}\label{appendix:shock_params_frame}
\subsection{Shock parameters and their estimation}\label{subsec:app_shock_params}
The upstream plasma conditions play a key role in determining the properties of collisionless shocks \citep{Burgess2015}. One of the most important geometric parameters is the angle between the shock normal and the upstream magnetic field, $\theta_{Bn}$. When $\theta_{Bn}$ approaches $90^\circ$, the magnetic field lies nearly tangent to the shock surface and the shock is classified as quasi-perpendicular. Smaller values of $\theta_{Bn}$ correspond to quasi-parallel configurations.

Shock dynamics are also characterized by several key dimensionless parameters. In particular, the Alfvénic and fast magnetosonic Mach numbers are defined as $M_A = v_{sh}/v_A$ and $M_{fms} = v_{sh}/c_{fms}$, where $v_{sh}$ is the shock speed in the upstream frame and $v_A$ and $c_{fms}$ are the upstream Alfvén and fast magnetosonic speeds. Additional quantities used to describe the shock include the upstream plasma beta ($\beta = v_{th}^2/v_A^2$) and the compression ratios of density and magnetic field across the shock transition, $r_G = n_d/n_u$ and $r_B = B_d/B_u$, where the subscripts $u$ and $d$ denote upstream and downstream quantities.

Determining these parameters from single-spacecraft observations is inherently challenging because shocks are three-dimensional and often highly variable structures \citep[e.g.][]{Koval2008}. In this work, shock normals are estimated using the Mixed-Mode~3 (MX3) method, while the shock speed is derived along the normal direction by enforcing mass flux conservation in the spacecraft frame. A detailed description of this technique can be found in \citet{Paschmann2000}. Once the shock normal and speed are obtained, additional quantities such as Mach numbers and compression ratios are derived from the upstream and downstream plasma measurements.

An important source of uncertainty in this procedure arises from the choice of upstream and downstream averaging intervals. Because the plasma environment around shocks is often highly structured, the inferred shock parameters can depend on the selected time windows \citep{Paschmann2000}. To account for this effect, we systematically vary the averaging intervals and compute distributions of parameter estimates following the approach described in \citet{Trotta2022b}. All parameters reported in this work are obtained using this method, implemented in the publicly available \texttt{SerPyShock} package\footnote{\url{https://github.com/trottadom/SerPyShock}}
 developed within the SERPENTINE project. In this work, we varied the averaging windows ranging from 30~s to 1~min, to capture the local regime of the shock front.

\subsection{Shock frames}

In the Solar Orbiter datasets, all vector quantities are expressed in spacecraft-centered Radial–Tangential–Normal (RTN) coordinates~\citep{Franz2002}, where $R$ points radially away from the Sun, $T$ is approximately along the orbital direction, and $N$ completes the right-handed system and is roughly northward with respect to the solar rotation axis.

For the analysis shown in panels f) of Figure~\ref{fig:overview}, plasma velocities are also expressed in the shock normal incidence frame (NIF)~\citep{Burgess2015}. In this coordinate system, the first axis is aligned with the shock normal $\hat{n}_{shock}$, while the two perpendicular directions are defined relative to the upstream magnetic field. Specifically, we construct $\hat{t}_2 = (\hat{n}_{shock} \times \mathbf{B}_u)/|\hat{n}_{shock} \times \mathbf{B}_u|$ and $\hat{t}_1 = \hat{t}_2 \times \hat{n}_{shock}$, where $\mathbf{B}_u$ is the average upstream magnetic field and $\hat{n}_{shock}$ is obtained from the MX3 method (see Section~\ref{subsec:app_shock_params}). This orthogonal basis defines the NIF coordinate system. In this frame, the upstream plasma velocity projected along the shock normal, $V_{NIF,0}$, is negative, indicating flow directed toward the shock. For further details, see~\citet{Dimmock2023}.

\section{Large scale SEP context}\label{appendix:seeds}
\begin{figure*}  
    \centering
   \includegraphics[width=0.99\textwidth]{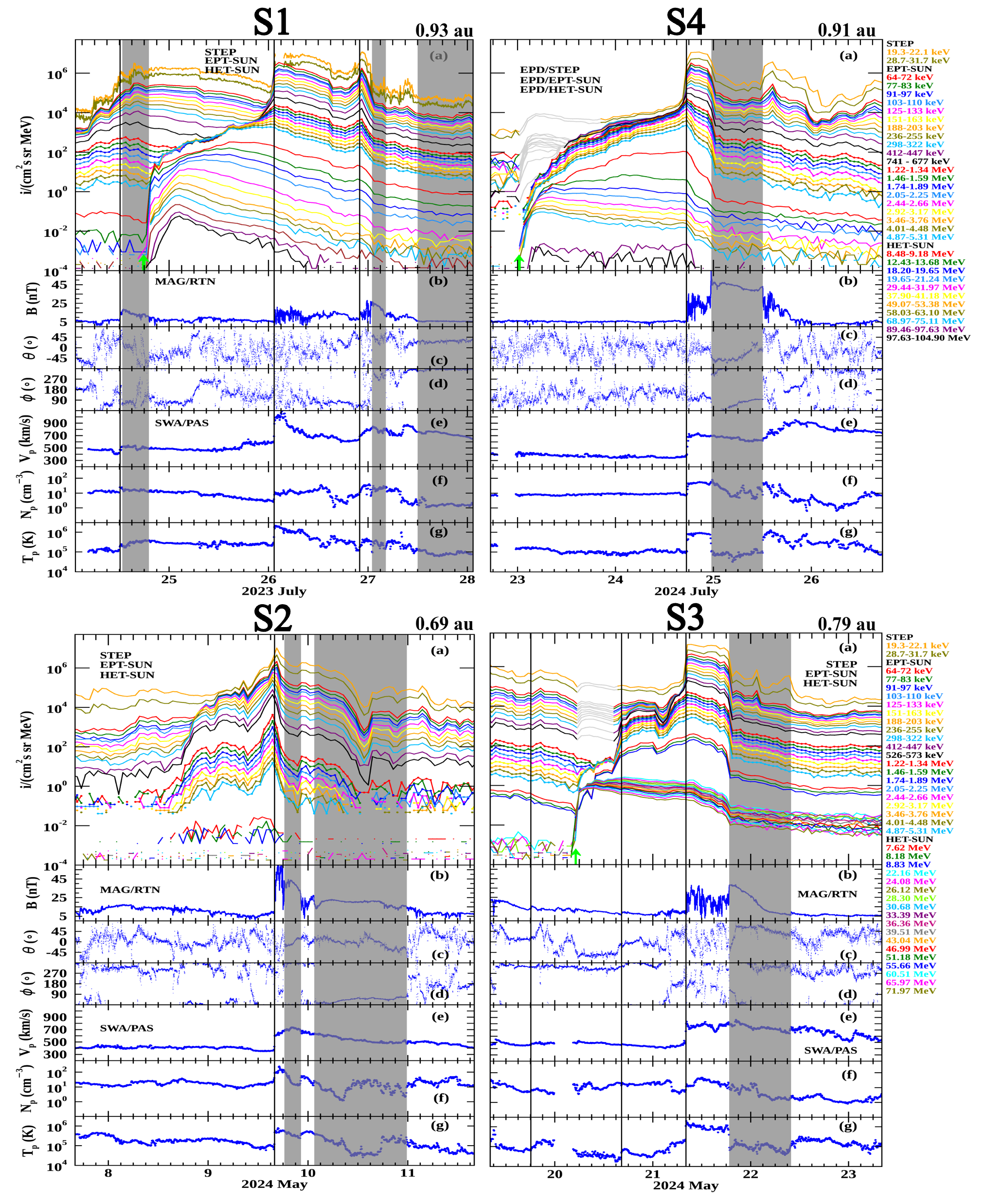}
   \caption{Large-scale context for the Shocks 1-4 (in counterclockwise order) as seen from Solar Orbiter. Each panel is center at the passage of the shock and covers two days prior and after the shock passage. Each subpanel shows (a) hourly averages of energetic particle intensities measured at selected energy channels of STEP and the sun apertures of EPT and HET; one-minute averages of magnetic field  (b) magnitude, and (c-d) angular direction in RTN coordinates; 10-minute averages of the solar wind proton (e) bulk speed, (f) density, and (g) temperature. The vertical solid lines indicate the passage of IP shocks and the shaded gray columns the passage of ICMEs as identified in helioforecast.space/icmecat \citep{moestl_2026}.}
    \label{fig:appendix_sep}%
\end{figure*}

We investigated the large-scale context of the energetic particle–dominated shock events. Figure~\ref{fig:appendix_sep} shows an overview of the in-situ Solar Orbiter observations for a period of two days prior and after the passage of the S1-4 (in counter-clokwise order). Panels (a) display energetic particle intensities at selected energy channels of STEP and the sunward-looking apertures of EPT and HET.  Panels (b-d) display the magnetic field magnitude and angular directions in RTN coordinates, and Panels (e-g) the solar wind proton bulk speed, $n_{p}$ and $T_{p}$. Note that during the periods of Shocks 2 and 3, the onboard EPD software used an outdated temperature configuration that affected the onboard HET calibration tables, making necessary a data recalibration to obtain particle intensities into physical units \citep{Lario_2026}.  The different energies during these two periods (indicated in Figure C.1)  have been used in computation of $P_{EP}$.

For S1, 3, and 4, the shock arrival is preceded by a strong injection of SEPs occurring at the time indicated by the green arrows. As a result, the upstream regions of these shocks contain a substantial seed population of pre-energized particles. For Shock 2, with a less energetic population of particles prior to the shock arrival,  a clear association with a solar eruption origin of both this injection of energetic particles and of the IP shock could not be unambiguously made \citep{Lario_2026}. Interestingly, this event also exhibits the shortest energetic particle–mediated precursor, although it displays the most pronounced intensity enhancement at the shock itself.

These observations suggest that the presence of pre-energized particles may contribute to the development of extended energetic particle–mediated precursors. However, the relationship between seed populations and shock-driven acceleration remains uncertain and is an active topic of investigation~\citep{Giacalone2026}.

\section{Wave foreshocks}\label{appendix:foreshock}

 \begin{figure*}  
    \centering
   \includegraphics[width=0.95\textwidth]{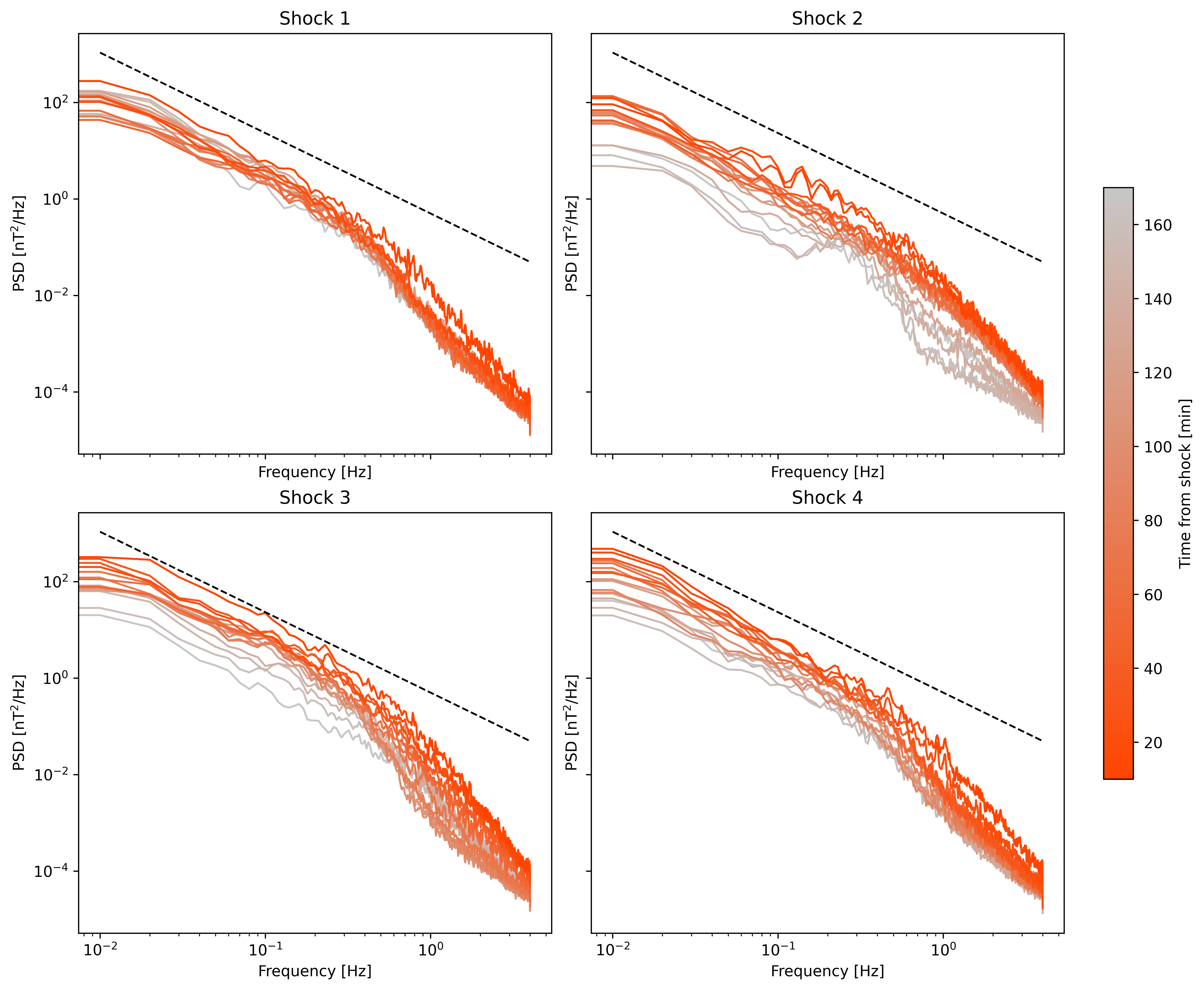}
   \caption{Magnetic field power spectral densities upstream of the four shock events. For each shock (panels 1–4), spectra are computed using sliding 20-minute windows shifted by 10 minutes and ordered by their time relative to the shock crossing, with darker colors corresponding to spectra closer to the shock. The dashed line shows a reference $f^{-5/3}$ slope.}
              \label{fig:appendix_foreshock}%
\end{figure*}

To investigate the presence of upstream wave activity associated with the energetic particle–mediated shocks, we analysed magnetic field fluctuations measured by the Solar Orbiter magnetometer (MAG). For each event, we examined a several hour interval surrounding the shock crossing, extending three hours upstream and 1.5 hours downstream of the shock.

Magnetic field power spectral densities (PSDs) were computed in successive sliding windows upstream of the shock. Each spectrum was obtained using a 20-minute interval with a 10-minute step, applying a Welch method to the magnetic field magnitude time series~\citep{Welch1967}. The resulting spectra were then ordered according to their temporal distance from the shock and displayed together to illustrate the evolution of magnetic fluctuations as the spacecraft approached the shock.

Figure~\ref{fig:appendix_foreshock} shows the PSDs for the four events. The color scale indicates the time relative to the shock arrival, allowing the progressive development of upstream fluctuations to be visualized. For reference, a $f^{-5/3}$ slope is shown, corresponding to the typical inertial-range scaling of solar wind turbulence~\citep{BrunoCarboneReview}. Deviations from this background spectrum and enhanced power at low frequencies indicate the presence of foreshock wave activity generated by particles streaming upstream of the shock.

From Fig.~\ref{fig:appendix_foreshock}, the clearest foreshock signatures are observed for Shocks~3 and~4. These events also exhibit the longest energetic particle–mediated precursors and correspond to the lowest values of $\theta_{Bn}$ in our sample. This behaviour is consistent with the expected dependence of foreshock wave activity on shock geometry, as quasi-parallel shocks allow particles to propagate further upstream and generate enhanced magnetic fluctuations. Albeit comparing IP foreshocks to the much better understood Earth's foreshock is an area of active research~\citep{Trotta2023a,Raptis2026}, we conclude that the  foreshock signatures shown in Figure~\ref{fig:appendix_foreshock} are not unique to the energetic particle–mediated regime, but are a natural consequence of particle reflection and streaming upstream of quasi-parallel shocks.

\section{Acknowledgements}\label{appendix:ack}
DT acknowledges support through the European Space Agency (ESA) Research Fellowship in Space Science. We acknowledge support from ESA through the Science Faculty - Funding reference ESA-SCI-E-LE-170. OP acknowledges support from ESA through the Archival Research Visitor Programme. P. M. acknowledges the partial support by NASA HGIO grant 80NSSC23K0419 and the NSF SHINE grant 2401162. OP is supported by the FIS2 Starting Grant FIS-2023-00246 “PhAse-sPAce cOmplexity in turbulent nearly-reversible plasmas (PAPAO)” (CUP B53C24009610001) funded by the Italian Ministry of University and Research. Work at Kiel university is supported by DLR grant 50 OC 2002.

\end{appendix}
\end{document}